\pdfoutput=1

\documentclass[11pt,twoside,a4paper,cmspaper,final,collab]{cms-tdr}
\def\svnVersion{347389}\def\cmsCernNoTag{CERN-PH-EP/2015-295}\def\cmsCernDate{\today}\def\cmsMessage{Published in the European Physical Journal C as \href{http://dx.doi.org/10.1140/epjc/s10052-016-4156-z}{\doi{10.1140/epjc/s10052-016-4156-z.}}}

\begin{document}\cmsNoteHeader{SMP-14-004}

\hyphenation{had-ron-i-za-tion}
\hyphenation{cal-or-i-me-ter}
\hyphenation{de-vices}
\RCS$Revision: 347334 $
\RCS$HeadURL: svn+ssh://svn.cern.ch/reps/tdr2/papers/SMP-14-004/trunk/SMP-14-004.tex $
\RCS$Id: SMP-14-004.tex 347334 2016-06-15 15:49:55Z jyhan $

\newlength\cmsFigWidth
\ifthenelse{\boolean{cms@external}}{\setlength\cmsFigWidth{0.85\columnwidth}}{\setlength\cmsFigWidth{0.4\textwidth}}
\ifthenelse{\boolean{cms@external}}{\providecommand{\cmsUpperA}{top\xspace}}{\providecommand{\cmsUpperA}{upper left\xspace}}
\ifthenelse{\boolean{cms@external}}{\providecommand{\cmsUpperB}{middle\xspace}}{\providecommand{\cmsUpperB}{upper right\xspace}}
\cmsNoteHeader{SMP-14-004}

\providecommand{\afb}{\ensuremath{A_\mathrm{FB}}\xspace}
\providecommand{\afbM}{\ensuremath{A_\mathrm{FB}(M)}\xspace}
\providecommand{\NA}{\ensuremath{\text{---}}\xspace}
\title{Forward-backward asymmetry of Drell--Yan lepton pairs in pp collisions at $\sqrt{s} = 8$\TeV}

\date{\today}

\abstract{A measurement of the forward-backward asymmetry \afb of oppositely charged lepton pairs ($\mu\mu$ and $\Pe\Pe$)  produced via $\Z/\gamma^*$ boson exchange in pp collisions at $\sqrt{s} = 8$\TeV is presented.
The data sample corresponds to an integrated luminosity of 19.7\fbinv collected with the CMS detector at the LHC.
The  measurement of \afb is performed for dilepton masses between 40\GeV and 2\TeV and for dilepton rapidity up to 5. The \afb measurements as a function of dilepton mass and rapidity are compared with the standard model predictions.}

\hypersetup{
	pdfauthor={CMS Collaboration},
	pdftitle = {Forward-backward asymmetry of Drell-Yan lepton pairs in pp collisions at sqrt(s) = 8 TeV},
	pdfsubject = {CMS},
	pdfkeywords = {CMS, physics, FB asymmetry, Drell-Yan}
}

\maketitle

\section{Introduction} \label{sec:intro}

A forward-backward asymmetry \afb in the production of Drell--Yan lepton pairs arises from the presence of both vector and axial-vector couplings of electroweak bosons to fermions. For a given dilepton invariant mass
 $M$ the differential cross section at the parton level at leading order (LO) can be expressed as
\begin{equation}
	\label{eq:crosssection}
	\frac{\rd\sigma}{\rd(\cos\theta^{*})} = A (1 + \cos^{2} \theta^{*}) + B  \cos\theta^{*},
\end{equation}
where $\theta^{*}$ represents the emission angle of the negatively charged lepton relative to the quark momentum in the rest frame of the dilepton system, and  $A$ and $B$ are parameters that depend on $M$, the electroweak mixing angle $\theta_\PW$, and the weak isospin and charge of the incoming and outgoing fermions.
The \afb quantity is
\begin{equation}
	{A}_\mathrm{FB} = \frac{\sigma_\mathrm{F}-\sigma_\mathrm{B}}{\sigma_\mathrm{F}+\sigma_\mathrm{B}},
\end{equation}
where $\sigma_\mathrm{F}$ ($\sigma_\mathrm{B}$) is the total cross section for the forward (backward) events, defined by $\cos\theta^{*} > 0$~($\cos\theta^{*} < 0$).
\afb depends on $M $, quark flavor, and the electroweak mixing angle $\theta_W$. Near the Z boson mass peak  \afb  is close to zero because of the small value of the lepton vector coupling to Z bosons. Due to weak-electromagnetic interference, \afb is large and negative for  $M$ below the Z peak ($M < 80$\GeV) and  large and positive above the \Z peak  ($M > 110$\GeV).  Deviations from the SM predictions could result from the  presence of additional neutral gauge bosons~\cite{London:1986, Rosner:1987a, Cvetic:1995zs, Rosner:1996, Bodek:2001}, quark-lepton compositeness~\cite{Abe:1997b}, supersymmetric particles, or extra dimensions~\cite{Davoudiasl:2000}. Around the \Z peak, measurements of  \afb  can also be used to extract the effective weak mixing angle $\sin^2\theta^\text{eff}_{\text{lept}}(m_{\Z})$~\cite{ATLASAfb,Nann:sin2thetaW} as well as the \PQu and \PQd quark weak coupling~\cite{D0:2008, cdf-ee, cdf-mumu, Nann:sin2thetaW}.

To reduce the uncertainties due to the transverse momentum (\pt) of the incoming quarks, this measurement uses the Collins--Soper (CS) frame~\cite{Collins:1977}. In this frame, $\theta^{*}_{\mathrm{CS}}$ is defined as the angle between the negatively charged lepton momentum and the axis that bisects the angle between the quark momentum direction and the opposite direction to the antiquark momentum. In the laboratory frame, $\theta^{*}_{\mathrm{CS}}$ is calculated as
\begin{equation}
        \label{cos}
        \cos\theta^{*}_{\mathrm{CS}}=\frac{2(P^+_{1}P^-_{2}-P^-_{1}P^+_{2})}{\sqrt{Q^2(Q^2+Q_\mathrm{T}^2)}},
\end{equation}
where $Q$ and $Q_\mathrm{T}$ represent the four-momentum and the \pt of the dilepton system, respectively, while
$P_{1}$ ($P_{2}$) represents the four-momentum of $\ell^{-}$ ($\ell^{+}$) with $P^\pm_{i} = (E_{i}\pm P_{z,i})/\sqrt{2}$, and $E_{i}$ represents the energy of the lepton.

The production of lepton pairs arises mainly from the annihilation of valence quarks with sea antiquarks.
 At the LHC, the quark and antiquark directions are not known for each collision because both beams consist of protons. In general, however, the quark carries more momentum than the antiquark as the antiquark must originate from the parton sea. Therefore, on average, the dilepton system is boosted in the direction of the valence quark~\cite{Rosner:1987a, Fisher:1994pw, Dittmar:1997}.  In this paper, the positive axis is defined to be along the boost direction using the following transformation on an event-by-event basis:
\begin{equation}
	\cos\theta^{*}_{\mathrm{CS}}\to\frac{\abs{Q_{ z}}}{Q_{z}}\cos\theta^{*}_{\mathrm{CS}},
\end{equation}
 where $Q_z$ is the longitudinal momentum of the dilepton system.
 The fraction of events for which  the quark direction is the same as the direction of the boost depends on $M$ and increases with the absolute value of the dilepton rapidity $y = \frac{1}{2}\ln[(E+Q_z)/(E-Q_z)]$.

 \afb was previously measured by the CMS~\cite{Yazgan:afb_comb} and ATLAS~\cite{ATLASAfb} experiments using data samples collected at $\sqrt{s} = 7$\TeV.
The techniques used in this analysis are similar to those used in the previous
 CMS measurement at 7\TeV, and the rapidity range of this measurement is extended to $\abs{y}=5$ by including  electrons in the forward calorimeter.
Since large Z boson rapidities are better correlated with the direction of the valence quark, \afb is measured as a function of the invariant mass and the rapidity of Z boson.
The number of selected events at 8\TeV is
about a factor of 5 larger than the number of events at 7\TeV.
The larger data sample collected at 8\TeV extends the measurement
of \afb in the high-mass region where the number of events in the 7\TeV
samples was limited.

\section{The CMS detector}

The central feature of the CMS detector is a superconducting solenoid with a 6\unit{m} internal diameter that provides a magnetic field of 3.8\unit{T}.
Inside the solenoid are a silicon pixel and strip tracker, a lead tungstate crystal electromagnetic calorimeter (ECAL), and a brass
and scintillator hadron calorimeter, each composed of a barrel and two endcap sections.
Extensive forward calorimetry complements the coverage provided by the barrel and endcap calorimeters.
Outside the solenoid, gas-ionization detectors embedded in the steel flux-return yoke are used to measure muons.

Muons are measured in the pseudorapidity~\cite{Chatrchyan:2008zzk} range $\abs{\eta}<2.4$ using the silicon tracker and muon systems. The muon detectors are constructed
using three different technologies: drift tubes for $\abs{\eta}<1.2$, cathode strip chambers for $0.9<\abs{\eta}<2.4$,
and resistive plate chambers for $\abs{\eta}< 1.6$. Matching muons to tracks measured in the silicon tracker
results in a relative \pt resolution of 1.3--2.0\% in the barrel, and better than 6\% in the endcaps for muons with $20 < \pt < 100$\GeV~\cite{Chatrchyan:2012xi}.

Electrons are measured in the range $\abs{\eta}<2.5$ using both the tracking system and the ECAL.
The energy resolution for electrons produced in Z boson decays varies from 1.7\% in the barrel ($\abs{\eta}<1.48$) to 4.5\% in the endcap region ($\abs{\eta}>1.48$)~\cite{Khachatryan:2015hwa}.

The $\eta$ coverage of the CMS detector is extended up to $\abs{\eta}=5$  by the hadron forward (HF) calorimeters~\cite{HF_performance}.
 The HF is constructed from steel absorbers as shower initiators and quartz fibers as active material. Half of the fibers
 extend over the full depth of the detector (long fibers) while the other half does not cover the first 22~cm measured
 from the front face (short fibers). As the two sets of fibers are read out separately, electromagnetic showers can be distinguished from hadronic showers.
 Electrons in the HF are measured in the range $3 < \abs{\eta} < 5$.
 The energy resolution for HF electrons is ${\sim} 32\%$ at 50 \GeV and the angular resolution is
 up to 0.05 in $\eta$ and $\phi$.

The CMS experiment uses a two-level trigger system. The level-1 trigger, composed of
custom-designed processing hardware, selects events of interest based on information from
 the muon detectors and calorimeters~\cite{L1_trigger}.
The high-level trigger is software based, running a faster version of the offline reconstruction
code on the full detector information, including the tracker~\cite{HLT}.
A more detailed description of the CMS detector, together with a definition of the coordinate system used and the relevant kinematic variables,
 can be found in Ref.~\cite{Chatrchyan:2008zzk}.

\section{Data and Monte Carlo samples}

The analysis is performed using the pp collision data collected with the CMS detector in 2012 at a center-of-mass energy of 8\TeV. The total integrated luminosity for the entire data set amounts to 19.7\fbinv.

The simulated $\Z/\gamma^* \to \mu\mu$ and $\Z/\gamma^* \to \Pe\Pe$ signal samples are generated at next-to-leading order (NLO) based in perturbative QCD using \POWHEG~\cite{Nason:2004rx, Frixione:2007vw, Alioli:2008gx, Alioli:2010xd} with the NLO CT10 parton distribution functions (PDFs)~\cite{PhysRevD.82.074024}.
The parton showering and hadronization are simulated using the \PYTHIA v6.426~\cite{Sjostrand:2006za} generator with the Z2* tune~\cite{z2tune}.

The background processes, $\Z/\gamma^* \to \tau\tau$, \ttbar, $\PQt\PWm$ and $\PAQt\PWp$, are generated with \POWHEG, and the inclusive W production with \MADGRAPH~\cite{Alwall:2011}.
The backgrounds from WW, WZ, and ZZ production are generated using \PYTHIA v6.426.
The $\tau$ lepton decays in the background processes are simulated using \TAUOLA~\cite{Davidson:2010rw}.
For all processes, the detector response is simulated using a detailed description of the CMS detector based on the \GEANTfour package~\cite{Sulkimo:2003, Allison:2006}. \textsc{GFlash}~\cite{CMS-CR-2009-343} is used for the HF~\cite{CMSPtY}, and the event reconstruction is performed with the same algorithms used for the data.
 The data contain multiple proton-proton interactions per bunch crossing (pileup) with an average value of 21. A pileup reweighting procedure is applied to the Monte Carlo (MC) simulation so the pileup distribution matches the data.

\section{Event selection}
The inclusive dimuon events are selected by a trigger that requires two muons, the leading one with $\pt>17$\GeV and the second one with $\pt>8$\GeV.
 Muons are selected offline by the standard CMS muon identification~\cite{Chatrchyan:2012xi}, which requires at least
 one muon chamber hit in the global muon track fit, muon segments in at least two muon stations, at least one hit in the pixel detector,
 more than five inner tracker layers with hits, and a $\chi^{2}/\mathrm{dof}$ less than 10 for the global muon fit. The vertex with the highest $\pt$ sum for
 associated tracks is defined as the primary vertex. The distance between the muon candidate trajectories and the primary vertex
is required to be smaller than 2\unit{mm} in the transverse plane and smaller than 5\unit{mm} in the longitudinal direction.
 This requirement significantly reduces the background from cosmic ray muons. To remove muons produced during jet fragmentation,
 the fractional track isolation, $\sum \pt^{\text{trk}} / \pt^{\mu}$, is required to be smaller than 0.1,
 where the sum runs over all tracks originating from the primary vertex within a cone of $\Delta R= \sqrt{\smash[b]{(\Delta\eta)^2 + (\Delta\phi)^2}}<0.3$ around each of the identified muons.
 Furthermore, each selected muon is required to have $\pt>20$\GeV and $\abs{\eta}<2.4$.

The inclusive dielectron events include electrons that are produced in an extended lepton pseudorapidity range, $\abs{\eta}<5$.
The events with dilepton rapidity $\abs{y}<2.4$ are selected by triggers requiring either two central electrons, $\abs{\eta}<2.4$, with $\pt>17$ and $>8$\GeV.
In the analysis, the central electron candidates are required to have $\pt>20$\GeV, have opposite charges, and to pass
 tight electron identification and isolation requirements~\cite{Khachatryan:2015hwa}.
The particle-flow (PF) event reconstruction~\cite{PFref1,PFref2} consists of reconstructing and identifying each single particle with an optimized combination of all subdetector information. In this process, the identification of the particle type (photon, electron, muon, charged hadron, or neutral hadron) plays an important role in the determination of the particle direction and energy.
 The fractional PF isolation, $\sum \pt^{\mathrm{PF}} / \pt^{\Pe}$, is required to be smaller than 0.1.
 The isolation variable is calculated from the energy sum over all PF candidates within a cone of size 0.3 around each of the identified electrons.
This sample is used to perform the analysis for the dilepton rapidity, $\abs{y}<2.4$.

For the events with dilepton rapidity $2.4 < \abs{y} < 5$, one central ($\abs{\eta}<2.4$) and one forward electron ($3<\abs{\eta}<5$) are used requiring one isolated central electron trigger with $\pt>27$\GeV.
In this case, the central (forward) electron candidate is required to have $\pt > 30\,(20)$\GeV,
as well as to pass stringent electron identification and isolation requirements (forward electron identification criteria).
Since the $2.4<\abs{\eta}<3$ region is outside the tracker acceptance, the particle flow variables cannot be defined in this region, and are therefore not considered in the analysis.

 Forward electron identification requires an isolated energy deposition in the core of the electron cluster~\cite{CMSPtY}.
 To reduce the contribution from jet background in the forward region, both electrons are required to be on the same side
 of the detector ($\eta_{\Pe_{1}} \, \eta_{\Pe_{2}} >0$) and almost back-to-back in azimuth ($\abs{\Delta\phi(\Pe_{1},\Pe_{2})}> 2\pi/3$).
Because the forward electrons do not have charge information, no oppositely-charged requirement is applied.

After the event selection, about 8 million $\mu\mu$ and 4.3 million $\Pe\Pe$ events remain with $\abs{y}<2.4$, and 0.5 million $\Pe\Pe$ events with $2.4<\abs{y}<5$.

\section{Simulation corrections}
Scale factors are derived and applied to the simulated MC events to account for differences of detector performance between data and the MC simulation.
The efficiencies for the trigger, lepton identification, and lepton isolation are measured using a ``tag-and-probe'' method~\cite{Khachatryan:2010,Chatrchyan:2012xi} for both data and simulation. For the muon channel, the trigger efficiency is measured as a function of  $\eta$ only because the $\pt$ dependence is small for $\pt>20$\GeV, while in the electron channel the efficiency is measured as a function of $E_{\mathrm T}$ and $\eta$. Similarly, the identification and isolation efficiencies for the muons and central electrons are measured in data and simulation as a function of $\pt$ and $\eta$.
The difference in trigger efficiency between data and simulation is 1 to 4\% for the muon channel, depending on the $\eta$ region, and less than 1\% for the electron channel. The differences in the muon identification and isolation efficiencies are less than 1\%.  For central electrons the absolute difference is at the 5\% level in the barrel and increases to 12\% in the endcaps.

 For forward electrons, the identification efficiency is measured as a function of $\ET$ and $\eta$.  We observe a 9 to 18\% difference in the identification efficiency between data and MC simulation.
The simulation is scaled using these factors to reproduce the data.
Forward electrons require additional corrections in \textsc{GFlash} simulation in order to match the $\eta$ distribution of the data.
Furthermore, a global normalization factor of $0.6 \pm 0.3$ is applied to  account for the data/simulation difference in the event yields in HF. Its effect is negligible in the \afbM measurement.

The muon momentum and electron energy scales are affected by detector misalignment and imperfect calibration, which cause a degradation in the energy measurements and the measurement of \afb. Such effects are accounted for by additional momentum and energy corrections, which are applied to muons and electrons in both data and simulation.
It has been shown~\cite{Chatrchyan:2012xi} that the primary cause of the bias in the reconstructed muon momentum is the misalignment of the tracking system.
To remove this bias, a muon momentum correction extracted as a function of the muon charge, $\theta$, and $\phi$~\cite{muscale} is applied for both data and MC events. The overall muon momentum corrections for muons with $\pt>20$\GeV are measured with a precision of better than 0.04\%.

For central electrons, an ECAL energy scale correction is applied. The overall energy scale for electrons with $7< \pt<70$\GeV is measured with a precision better than 0.3\%~\cite{Khachatryan:2015hwa}. To match the electron energy resolutions in data, additional smearing is applied to the energy of central electrons in the MC simulation.
For forward electrons, the predicted energy of the forward electron is calculated using Z boson mass, the energy of the central electron, and the angular positions ($\eta$ and $\phi$) of central and forward electrons.
The residual energy correction for forward electrons as a function of $\ET$ is determined from the average of the difference between the reconstructed energy and the predicted energy.
The corrections are applied in data and simulation as a function of the electron $\ET$ and
 range between $-18\%$ and $+12\%$.
The energy resolution of the forward electron in the MC simulation is also tuned to match the data.

\section{Backgrounds}

The main sources of background at low dilepton mass are  $\Z/\gamma^{*} \to \tau\tau$  events and QCD dijet events.
At high mass, the main background comes from t$\bar{\text{t}}$ events.
The diboson (WW, WZ, ZZ) and inclusive W background contributions are small.
The background contributions are estimated versus $M$ and $\abs{y}$ for forward and backward events separately.
Different techniques are used for estimating background contributions in the muon and electron channels.

 The dijet background for both muon and electron channels is estimated with data using control samples.
The muon channel  uses same-sign dimuon events, which mostly originate from dijets.
The number of same-sign events after the final event selection is used to estimate the number
of opposite-sign dimuons that originate from dijets.
 The contribution from the diboson process
 is subtracted in the same-sign events using MC simulation.

For the electron channel, a fitting method is used to estimate the dijet background.
 The  kinematic distributions of the $\Pe\Pe$ events in $M$ and $\abs{y}$ are fitted with a sum of signal
 and background templates to determine the dijet component.   A signal template is extracted from
the  $\Z/\gamma^{*} \to \Pe\Pe$ MC sample.
A background template is obtained by applying a reverse isolation requirement on the central electron in data.
The signal and non-QCD background contributions, which are small, are subtracted from this nonisolated electron sample using simulation.

In the muon channel, events selected with an $\Pe\mu$ lepton pair are used to determine
the backgrounds from $\Z/\gamma^{*} \to \tau\tau$, $\ttbar$, W+jets, $\PQt\PW$, and $\PAQt\PW$ processes.
The overall rate for $\mu\mu$ background events from these sources is proportional to the number of observed $\Pe\mu$ events.
Here the MC simulation is used only to calculate the ratio of $\mu\mu$ events to $\Pe\mu$ events.
The background rate extracted with this method is in agreement with MC simulations.
Therefore, in the  electron analysis these backgrounds are modelled using MC simulations.
The  cross sections are normalized to next-to-next-to-leading-order \FEWZ predictions~\cite{FEWZ}.
Also, the diboson backgrounds are estimated using MC simulation for both the muon and electron channels.

\begin{figure}[!hbt]
\centering
\includegraphics[width=0.48\textwidth]{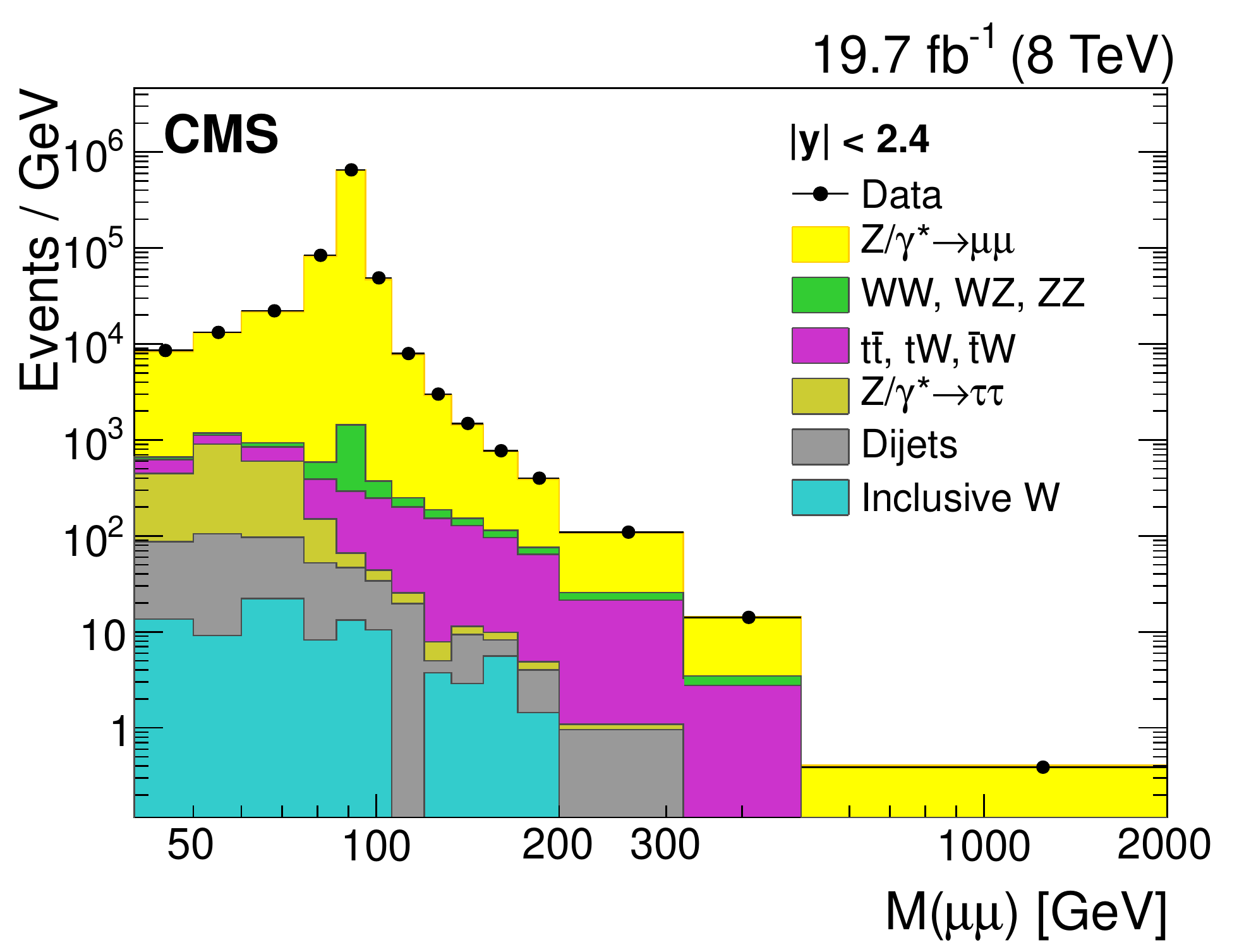}
\includegraphics[width=0.48\textwidth]{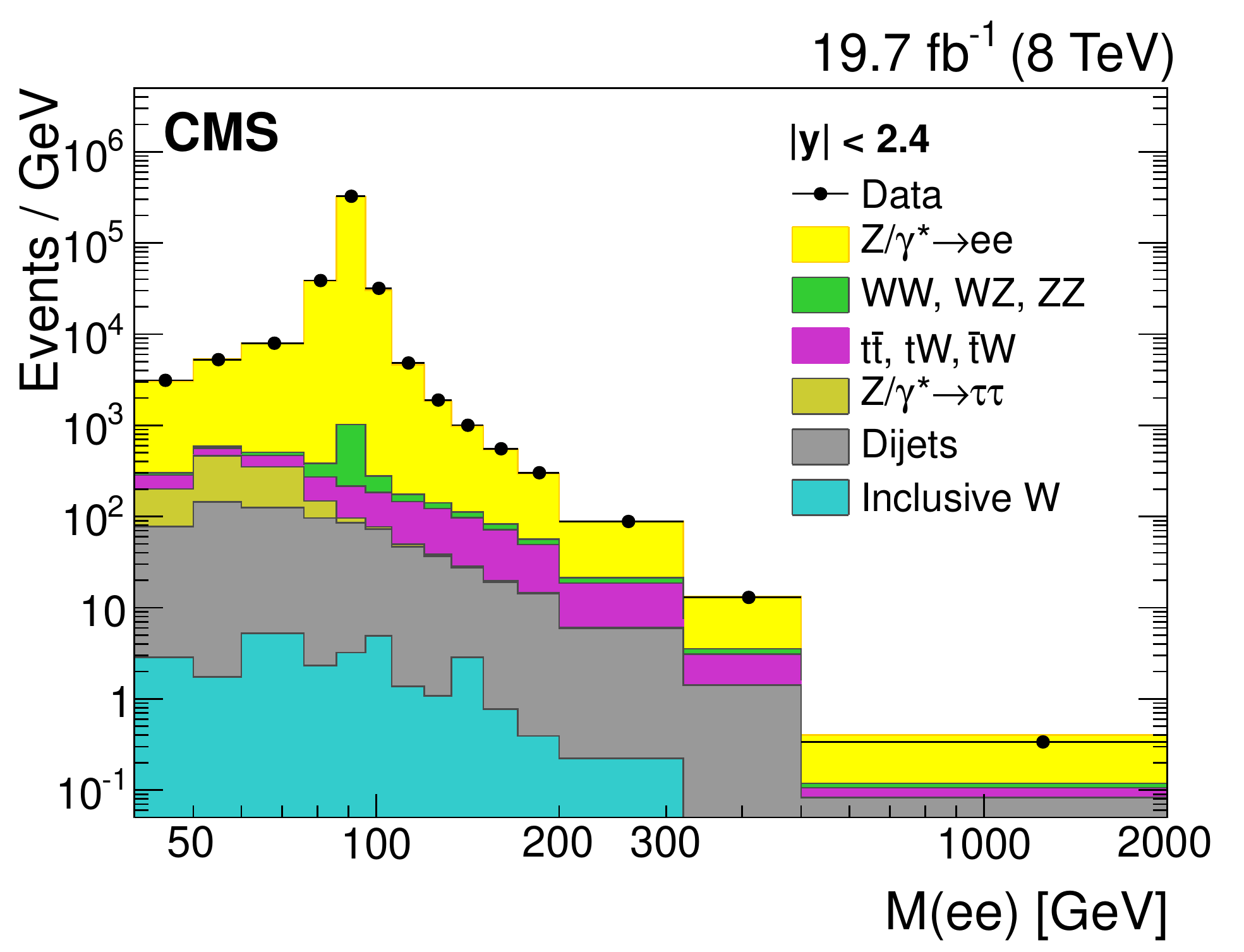}
\includegraphics[width=0.48\textwidth]{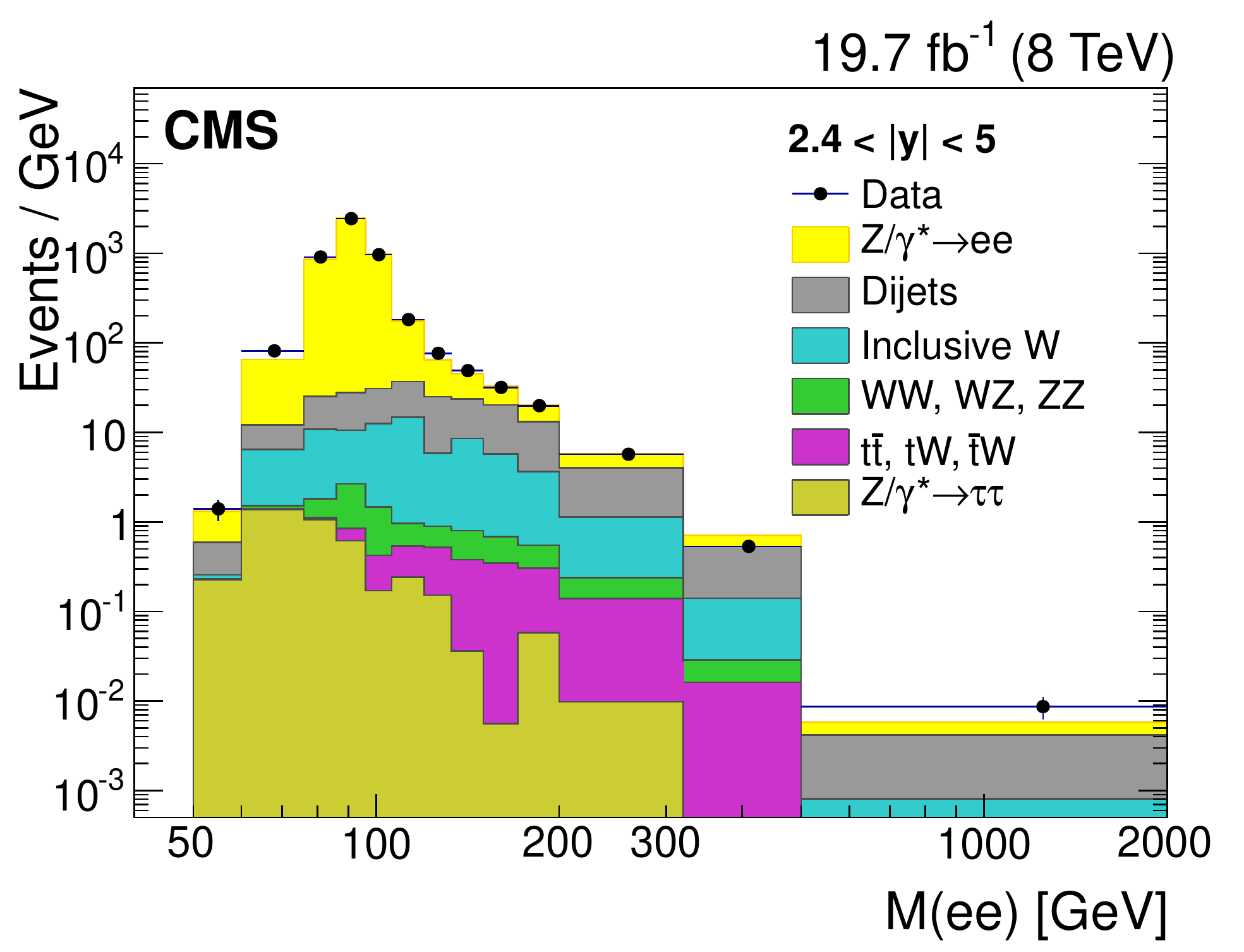}
\caption{The invariant mass distributions for $\mu\mu$ (\cmsUpperA), $\Pe\Pe$ (\cmsUpperB) events with $\abs{y}<2.4$, and $\Pe\Pe$ (bottom) events with $2.4<\abs{y}<5$. Only statistical uncertainties are shown. The stacked histograms represent the sum of the background contributions and the signal.
}
\label{fig:muon_bkg}

\end{figure}

The invariant mass distributions for $\mu\mu$ and  $\Pe\Pe$ events in two $\abs{y}$ ranges are shown in Fig.~\ref{fig:muon_bkg}, which also includes the MC predictions
 for both the signal and estimated background contributions.
The MC predictions are normalized using the cross section for each process and the integrated luminosity.

\section{Measurement of \texorpdfstring{\afb}{A[FB]}} \label{sec:unfolding}
\begin{figure*}[htbp]
\centering
\includegraphics[width=0.48\textwidth]{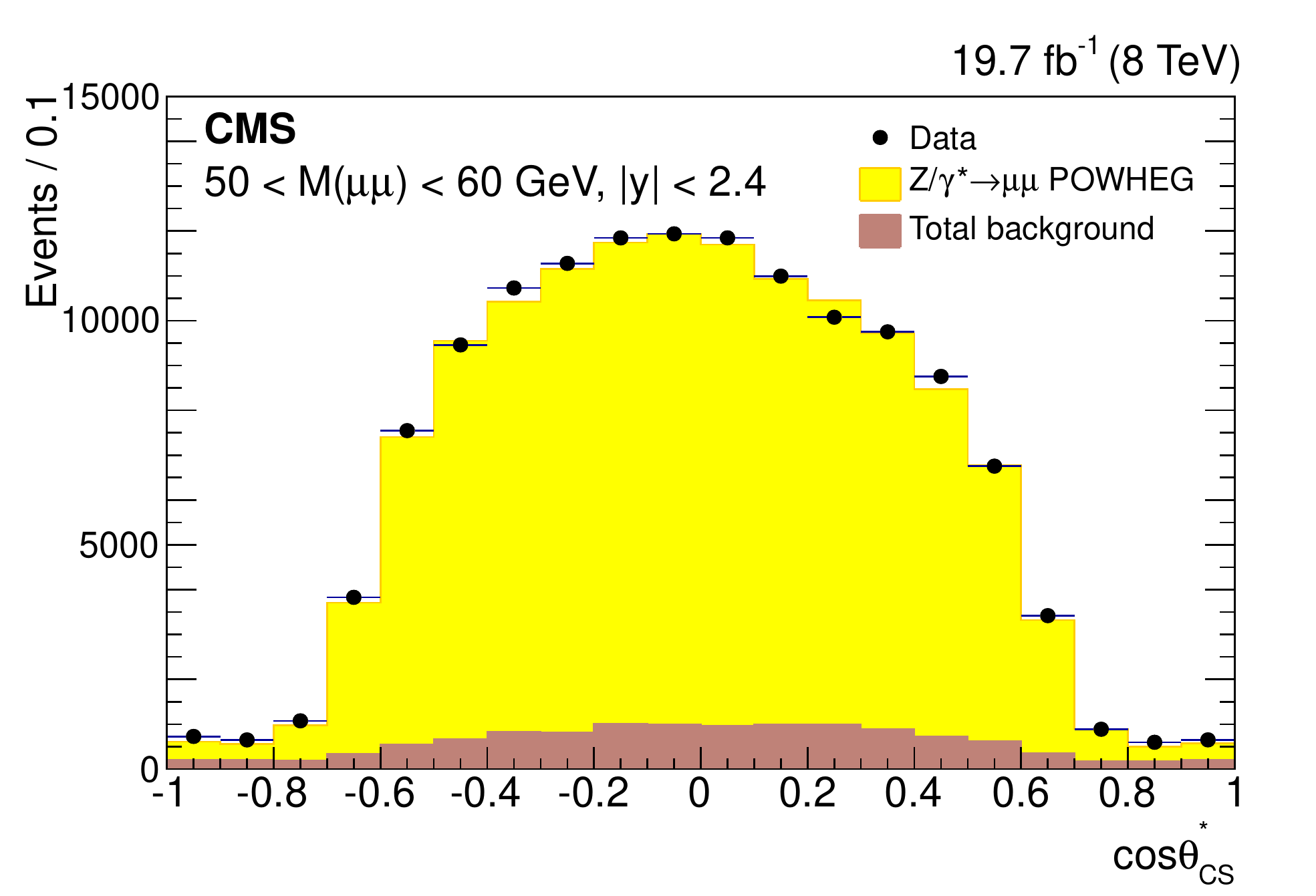}
\includegraphics[width=0.48\textwidth]{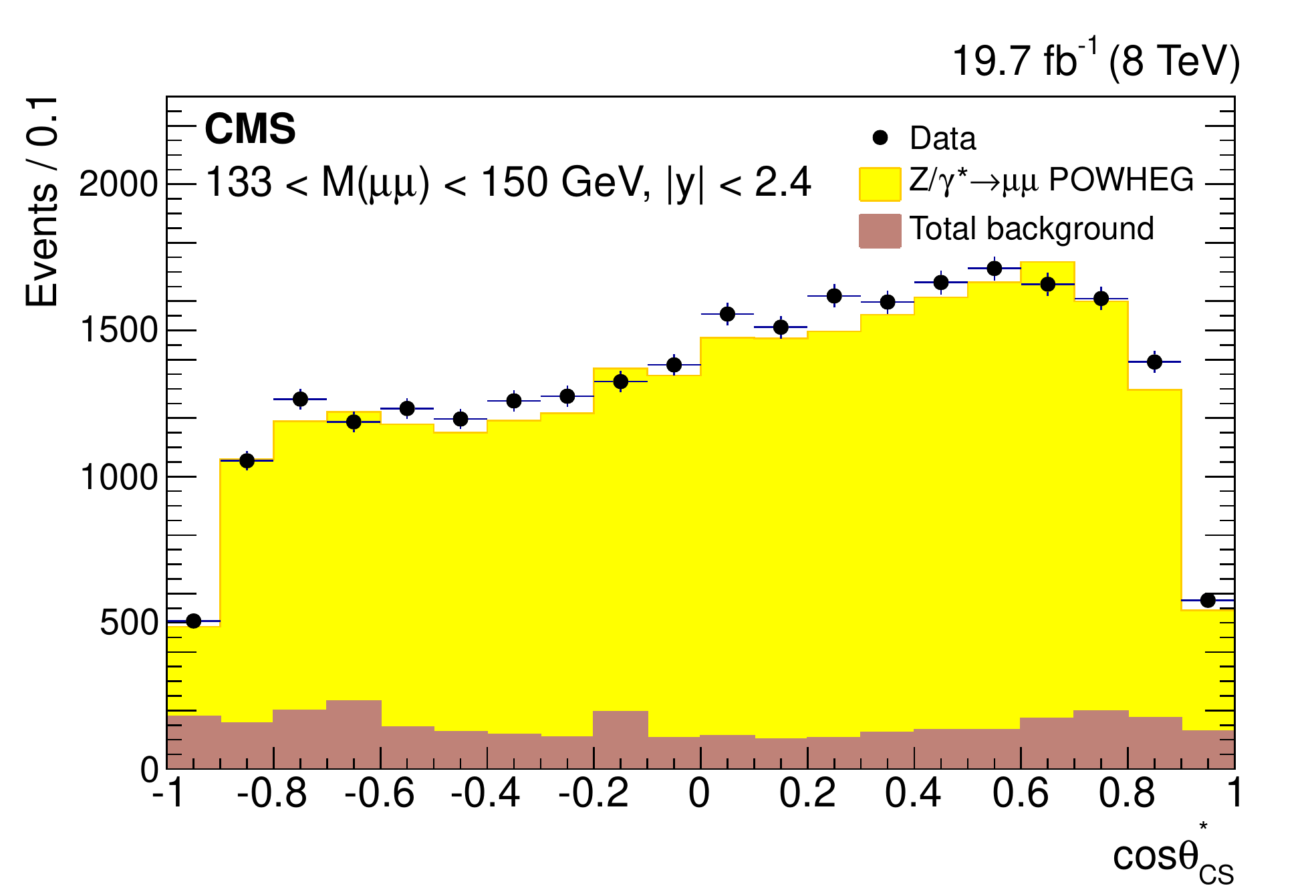}
\includegraphics[width=0.48\textwidth]{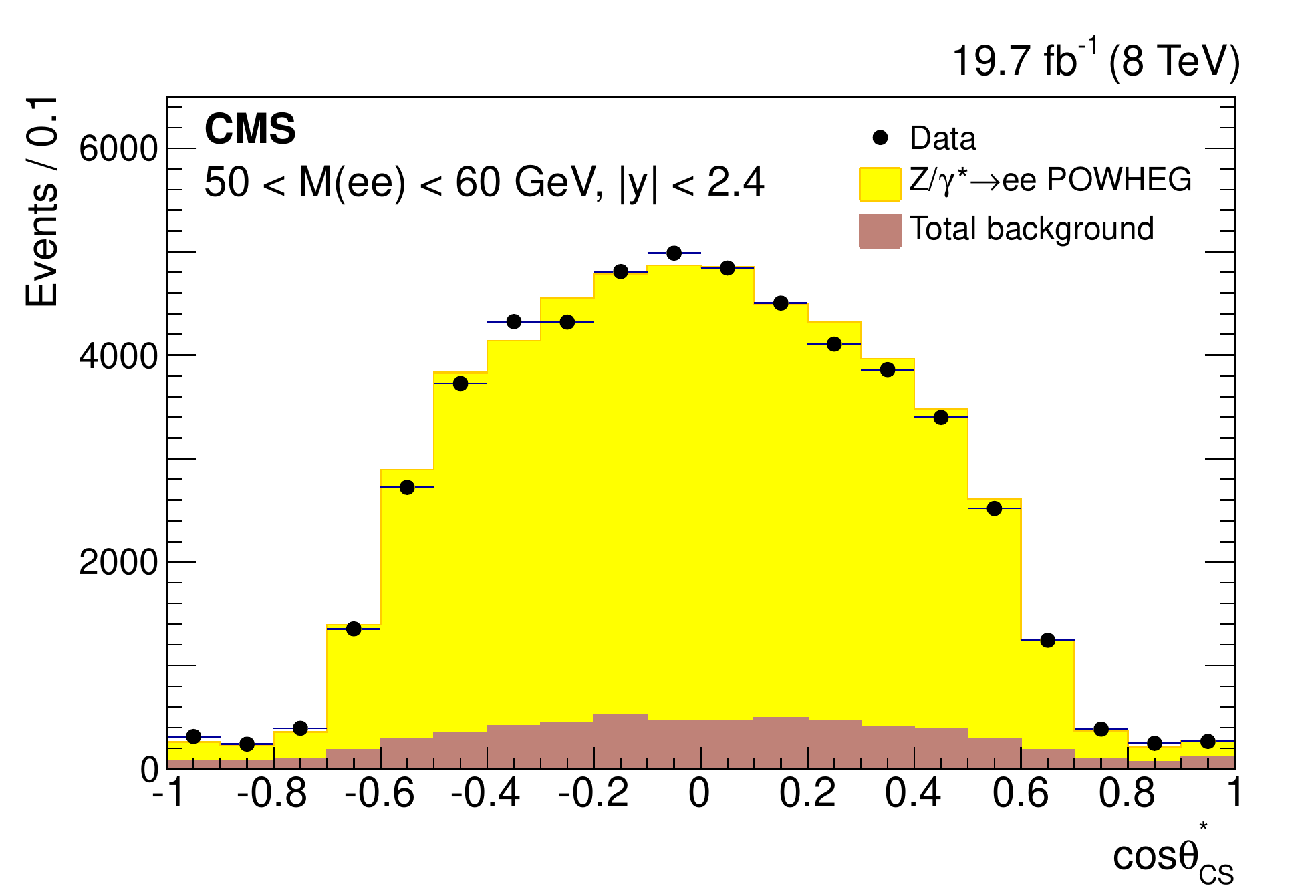}
\includegraphics[width=0.48\textwidth]{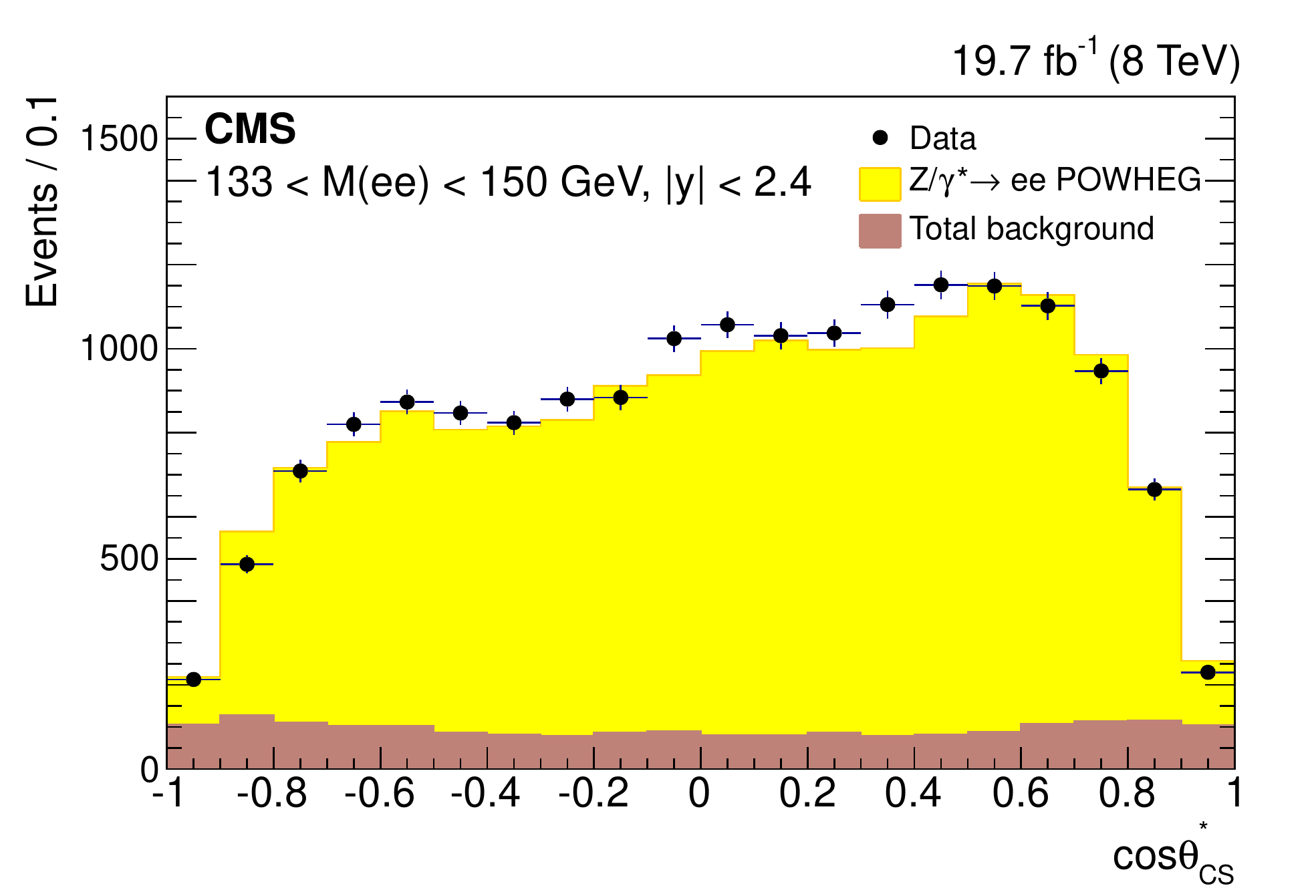}
\caption{The $\cos\theta^{*}_\mathrm{CS}$ distributions for $\mu\mu$ ($\Pe\Pe$) events
 are presented in the top (bottom) panels. Only statistical uncertainties are shown.
 The stacked histograms represent the sum of the background contribution and the signal.
 The plots on the left (right) panels correspond to events with dilepton invariant mass $50 < M < 60$\GeV ($133 < M < 150$\GeV). }
\label{fig:cos_dist}
\end{figure*}

The events are assigned to ``forward'' or ``backward'' regions as described in Section~\ref{sec:intro}.
\afb is measured using the selected dilepton events as a function of dilepton mass
 in five regions of absolute rapidity: 0--1, 1--1.25, 1.25--1.5, 1.5--2.4, and 2.4--5.
The most forward region has 7 mass bins, from 40 to 320\GeV, while the others have 14 mass bins,
 which extend up to 2\TeV.
The shape of the $\cos\theta^{*}_\mathrm{CS}$ distribution changes with the dilepton mass.
The top panels of Fig.~\ref{fig:cos_dist} show the reconstructed $\cos\theta^{*}_{\mathrm{CS}}$
 distributions for $\mu\mu$ events, with $\abs{y} < 2.4$. The
bottom panels show the  reconstructed $\cos\theta^{*}_\mathrm{CS}$ for
$\Pe\Pe$ events, with $\abs{y} < 2.4$.  The distributions are shown  for two representative mass bins.
The distributions for dilepton events at low mass ($50 < M < 60$\GeV) are shown in the left panels,
 and at high mass ($133 < M < 150$\GeV) in the right panels.
The MC predictions are normalized to the integrated luminosity of the data.

{\tolerance=600
The measured  \afb value is corrected for detector resolution, acceptance, efficiency, and the effect of final-state QED radiation (FSR)  using a two-dimensional iterative unfolding method based on Bayes' theorem~\cite{D'Agostini:1994zf, RooUnfold:2011}.
The \afb quantity is unfolded to account for  event migration  between mass bins and between positive and negative $\cos\theta^{*}_\mathrm{CS}$ region.
Since the ambiguity of the quark direction is more significant at low $\abs{y}$, the dilution of \afb is larger in the low $\abs{y}$ region.
\par}

\section{Systematic uncertainties} \label{sec:systematics}

The  largest experimental uncertainties originate from the background estimation,
 the electron energy correction, the muon momentum correction, and the unfolding procedure.
The dominant contribution to the background uncertainty is the statistical uncertainty
 in the background data control sample. The theoretical uncertainty of the cross section in
 the MC background samples also contributes to the systematic uncertainty
 in the estimation of the background.

After energy corrections to central electrons are applied,  we find that there is
a 0.4\%  offset in the  position of the Z peak between data and simulation
in the  barrel and a  0.5\% offset in  the endcaps. This difference is assigned as the systematic uncertainty
 in the central electron energy calibration.

 In order to estimate the uncertainty in the  energy calibration of forward electrons, the parametrized function of the correction factor is scaled up and down by its statistical uncertainty.
The difference in  \afb before and after changing the correction factor is assigned
 as a systematic uncertainty.

 The systematic uncertainty in the muon momentum correction is estimated with a similar approach.
 The muon momentum correction is scaled up and down by its statistical uncertainty and the difference in \afb
 resulting from the change of the muon momentum correction is assigned as systematic uncertainty.
We find that the  contributions of the  uncertainties in the efficiency scale factors
 (trigger, identification, and isolation) and in the pileup reweighting factors to the uncertainty in \afb are small.

 For forward HF electrons, the uncertainties in the electron $\eta$ correction and in the global normalization factor
contribute to the systematic uncertainty in \afb.
In addition, the energy calibration varies approximately 5\% between $+\eta$ and $-\eta$.
 To account for this asymmetric effect in the energy calibration,
 the \afb distribution is measured using one forward electron in $+\eta$ or $-\eta$, separately,
  along with one central electron and half of the difference in \afb is assigned as a systematic uncertainty.
 The systematic uncertainty varies from 0.005 to 0.03 as a function of dielectron invariant mass.

 The systematic uncertainty in the unfolding procedure is estimated using a closure test in simulation.
 Any residual shown in the closure test of the unfolding procedure is assigned as the systematic uncertainty.

The theoretical uncertainties which affect the detector acceptance originate from the uncertainties in  PDFs (CT10~\cite{PhysRevD.78.013004, PhysRevD.82.074024} and NNPDF
2.0~\cite{Ball:2010de}) and  from uncertainties in the FSR modeling~\cite{Khachatryan:2010xn}.

 The systematic uncertainty in \afb depends on the mass of the dilepton pair.
Table~\ref{tab:syst} gives the maximum value of this uncertainty from each source, for different regions of $\abs{y}$.

\begin{table*}[htb]
\centering
\topcaption{ The maximum  value of the  systematic uncertainty in  \afb as a function of $M$ from each source for different regions of $\abs{y}$.}
\label{tab:syst}
Muon channel \\[2ex]
\begin{tabular}{lrccc} \hline
 \multirow{ 2}{*}{Systematic uncertainty}                     &  \multicolumn{4}{ c}{$\abs{y}$ bins } \\ \cline{2-5}
 & \multicolumn{1}{c}{0--1} & 1--1.25 & 1.25--1.5 & 1.5--2.4 \\  \hline
Background &  0.062 & 0.080 & 0.209 & 0.051 \\
Momentum correction &  0.006 & 0.015 & 0.020 & 0.022  \\
Unfolding  &  0.001 & 0.003 & 0.004 & 0.003  \\
Pileup reweighting & 0.002 & 0.004 & 0.003 & 0.004  \\
Efficiency scale factors & $<$0.001 & 0.002 & 0.003 & 0.005  \\
PDFs & 0.001 & 0.004 & 0.008 & 0.047 \\
FSR & $<$0.001 & 0.001 & 0.001 & 0.002 \\  \hline
\end{tabular} \\[5ex]
Electron channel \\[2ex]
\begin{tabular}{lrrrrr} \hline
  \multirow{ 2}{*}{Systematic uncertainty}   &  \multicolumn{5}{ c}{$\abs{y}$ bins } \\ \cline{2-6}
 &  \multicolumn{1}{c}{0--1} & \multicolumn{1}{c}{1--1.25} & \multicolumn{1}{c}{1.25--1.5} & \multicolumn{1}{c}{1.5--2.4} & \multicolumn{1}{c}{2.4--5} \\ \hline
Background &  0.064 & 0.015 & 0.008 & 0.004 & 0.033\\
Energy correction &  0.011 & 0.015 & 0.012 & 0.012 & 0.123 \\
Unfolding  &  0.005 & 0.007 & 0.006 & 0.004 & 0.001 \\
Pileup reweighting & 0.003 & 0.002 & 0.002 & 0.001 & 0.007 \\
Efficiency scale factors & $<$0.001 & $<$0.001 & $<$0.001 & $<$0.001 & 0.008 \\
Forward $\eta$ scale factor & \multicolumn{1}{c}{\NA} & \multicolumn{1}{c}{\NA} & \multicolumn{1}{c}{\NA} & \multicolumn{1}{c}{\NA} & 0.002 \\
Forward $\eta$ asymmetry & \multicolumn{1}{c}{\NA} & \multicolumn{1}{c}{\NA} & \multicolumn{1}{c}{\NA} & \multicolumn{1}{c}{\NA} & 0.029 \\
Global normalization factor & \multicolumn{1}{c}{\NA} & \multicolumn{1}{c}{\NA} & \multicolumn{1}{c}{\NA} & \multicolumn{1}{c}{\NA}  & 0.060 \\
PDFs & 0.002 & 0.004 & 0.005 & 0.008 & 0.014 \\
FSR & $<$0.001 & 0.001 & 0.001 & 0.001 & 0.002 \\ \hline
\end{tabular}
\end{table*}

\section{Results}

\begin{figure*}[hbtp]
	\centering
		\includegraphics[width=0.48\textwidth]{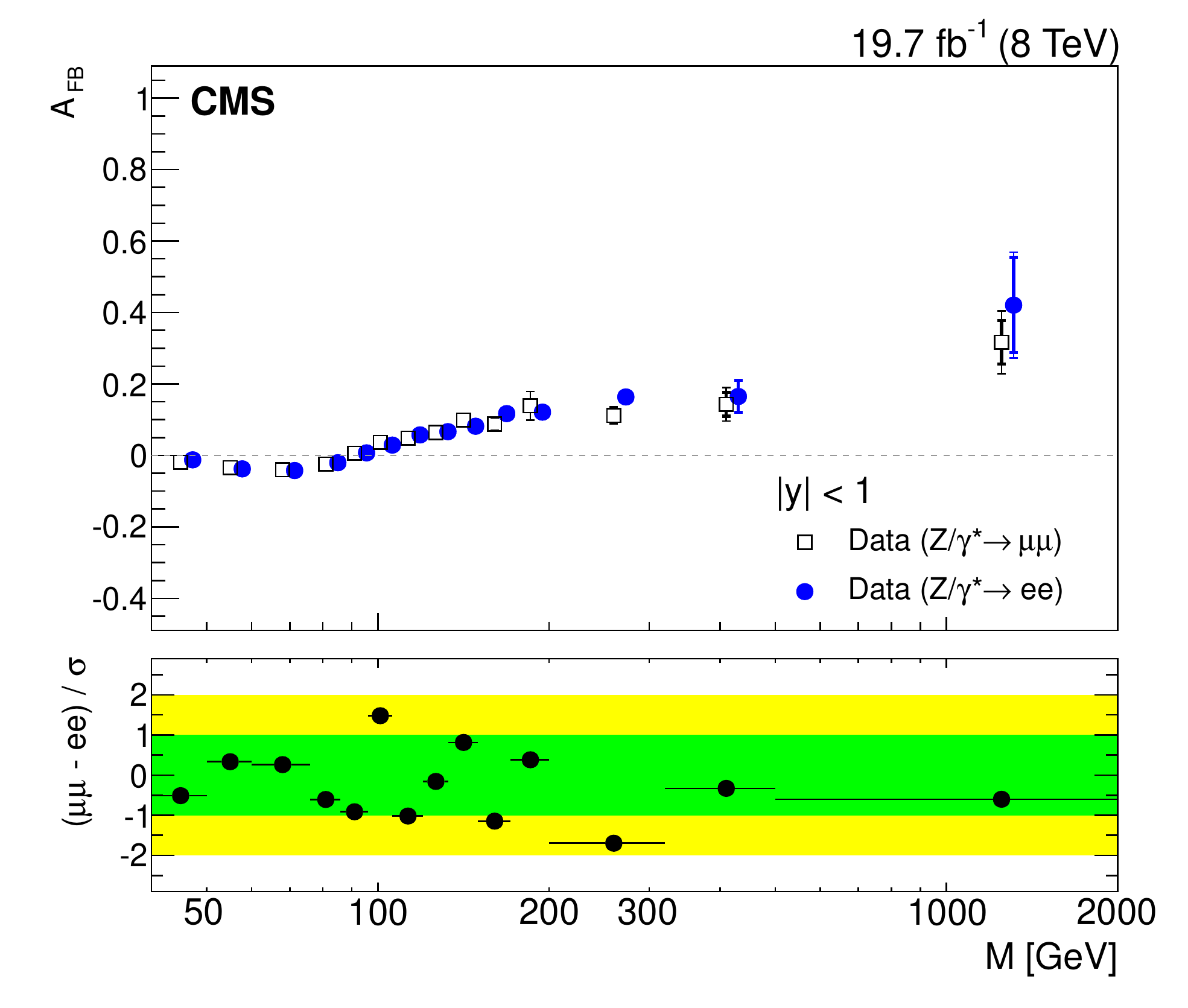}
		\includegraphics[width=0.48\textwidth]{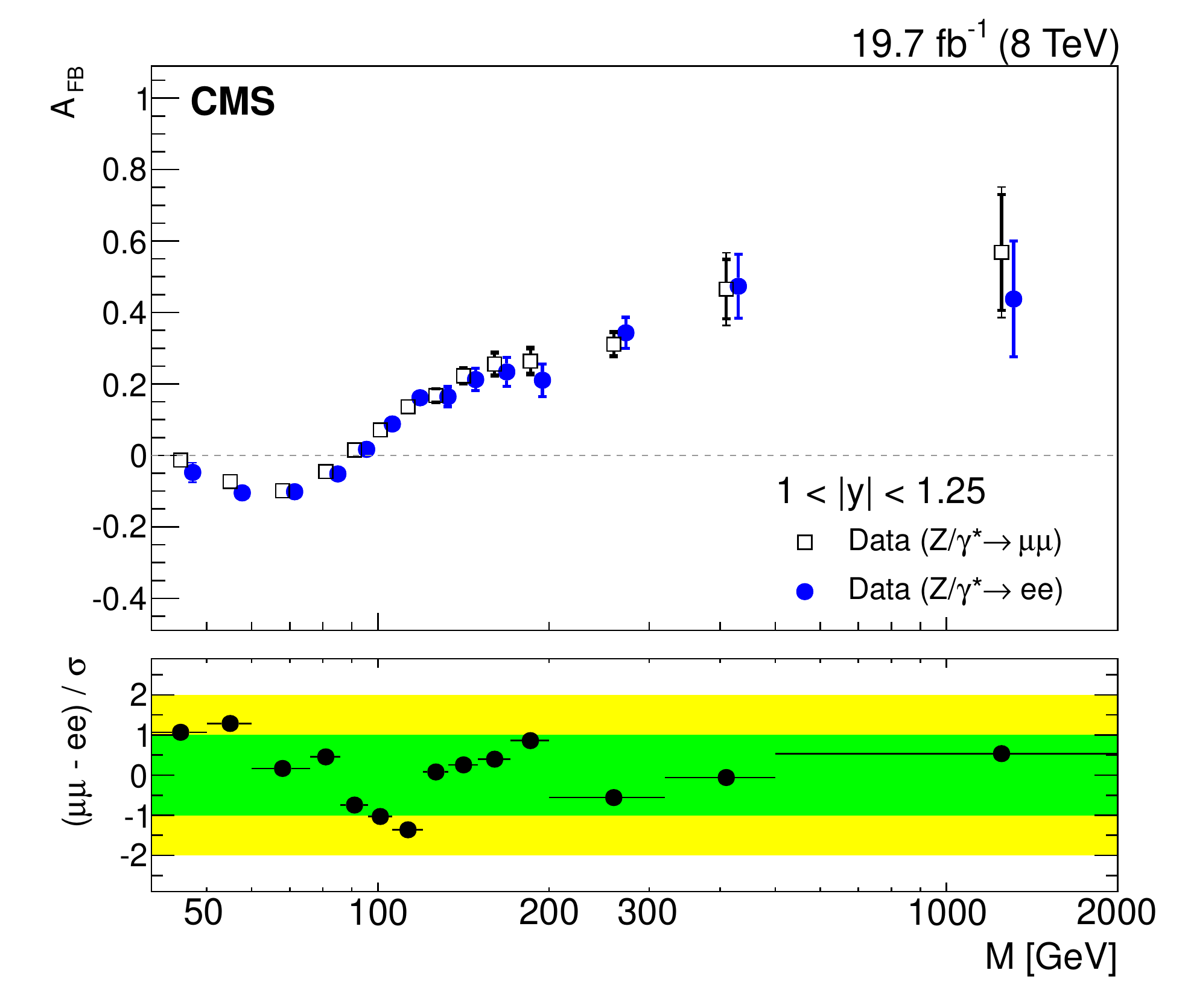}
		\includegraphics[width=0.48\textwidth]{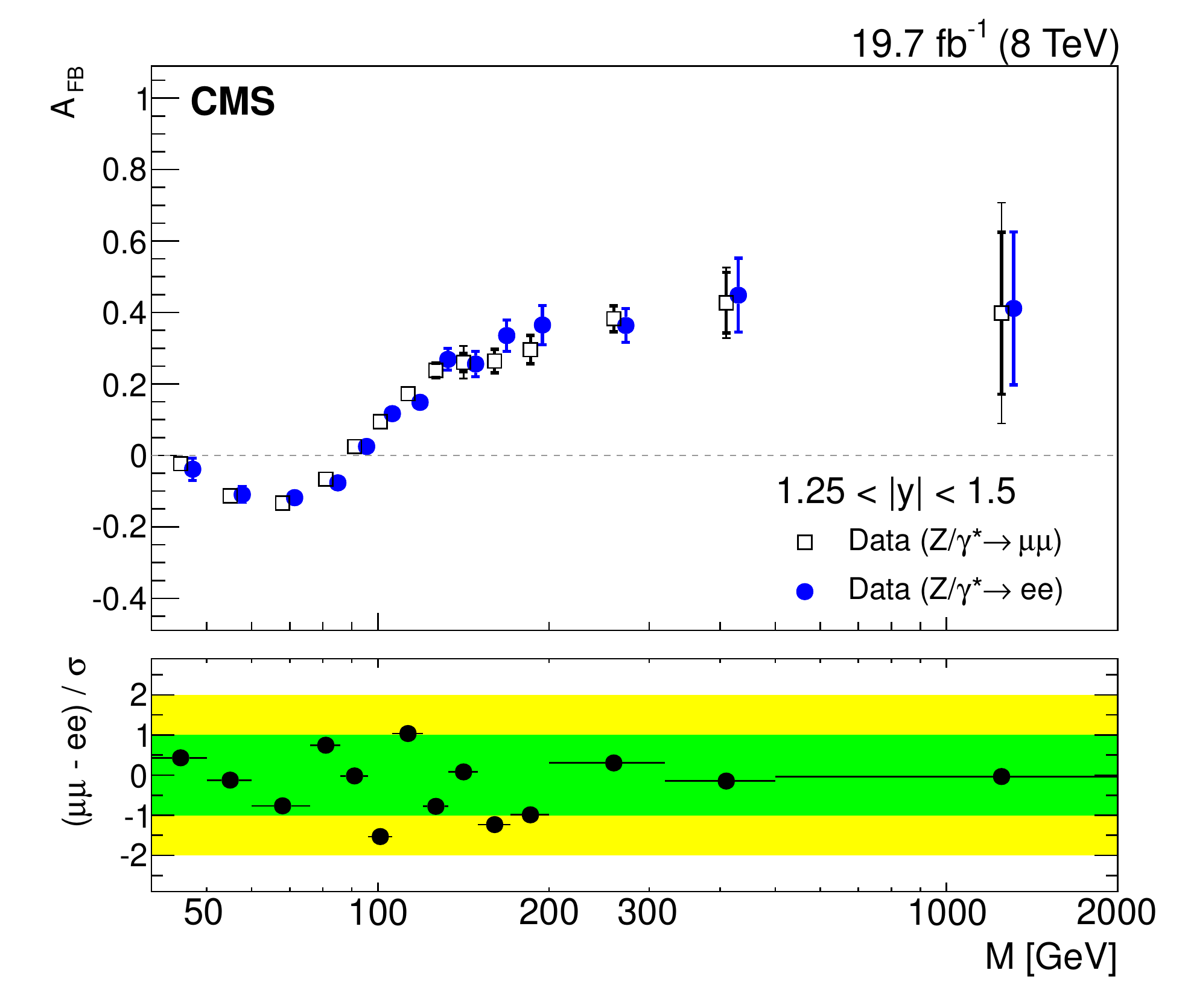}
		\includegraphics[width=0.48\textwidth]{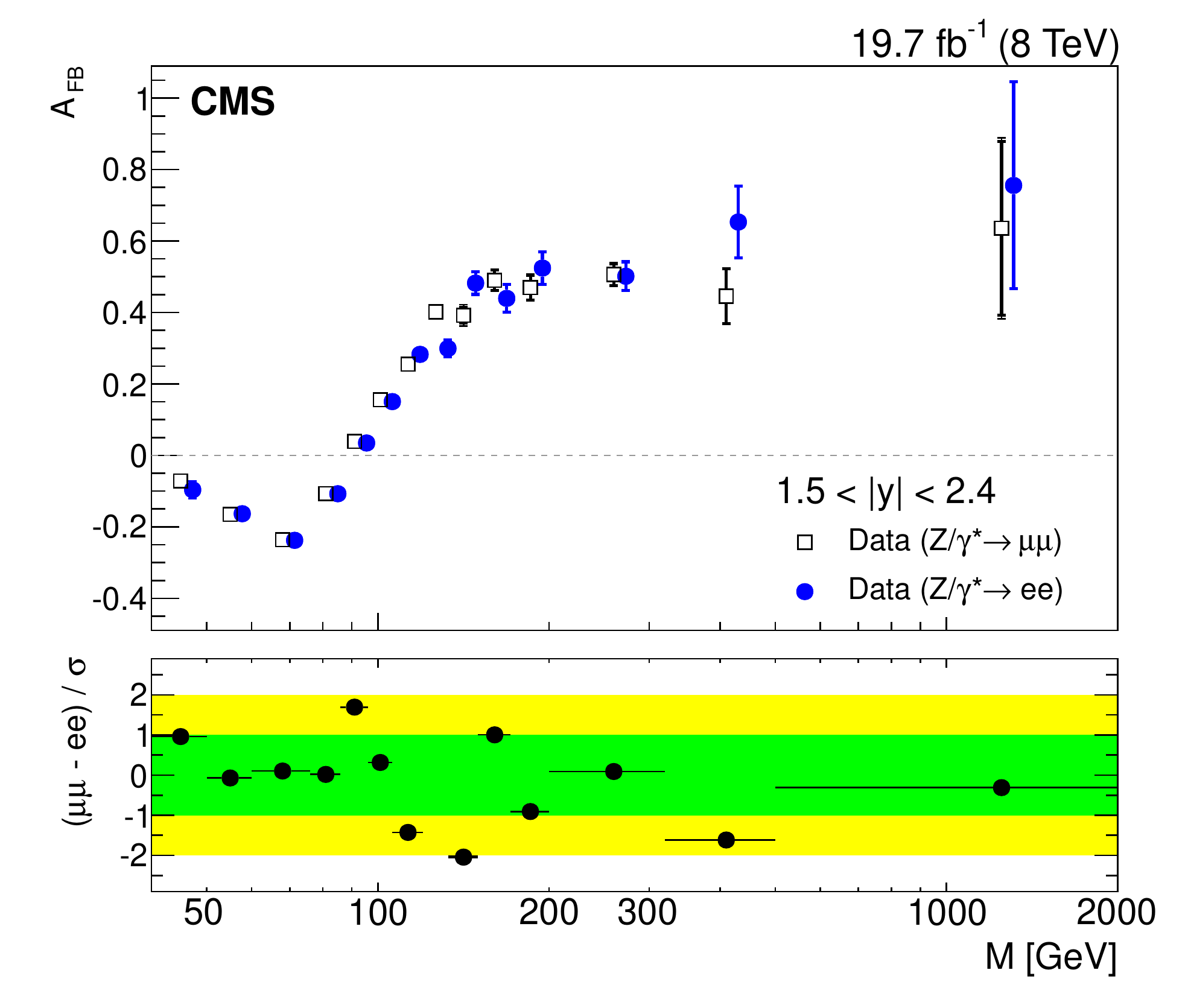}
		\caption{The unfolded \afb distributions for  muons (open squares) and  electrons (solid circles)  for the four central rapidity regions.
The statistical (thick vertical bar) and statistical plus systematics (thin vertical bar) uncertainties are presented.
The solid circles are shifted slightly to compare the result better. The lower panel in each plot shows the difference
 of the unfolded \afb in muons and electrons divided by the total uncertainty (stat. $\oplus$ syst.).
}
\label{fig:Afb_unf}	
\end{figure*}

{\tolerance=600
A comparison of the unfolded, background-subtracted \afbM distributions for $\mu\mu$ and $\Pe\Pe$ events in the four central rapidity regions
is shown in Fig.~\ref{fig:Afb_unf}. The statistical and systematic uncertainties are added in quadrature.
The measured  \afbM distributions agree for $\mu\mu$ and $\Pe\Pe$ events in all rapidity regions.
\par}

The unfolded  \afbM measurements for $\mu\mu$ and $\Pe\Pe$ events, within $\abs{y} < 2.4$, are combined under the assumption
that the  uncertainties  in the  muon and electron channels are uncorrelated.
Any effect of the correlation between  the $\mu\mu$ and $\Pe\Pe$ systematic uncertainties in the pileup correction, FSR modeling, and the normalization of MC simulations in the background estimation is found to have a negligible effect on the combination.

\begin{figure*}[htbp]
	\centering
		\includegraphics[width=0.48\textwidth]{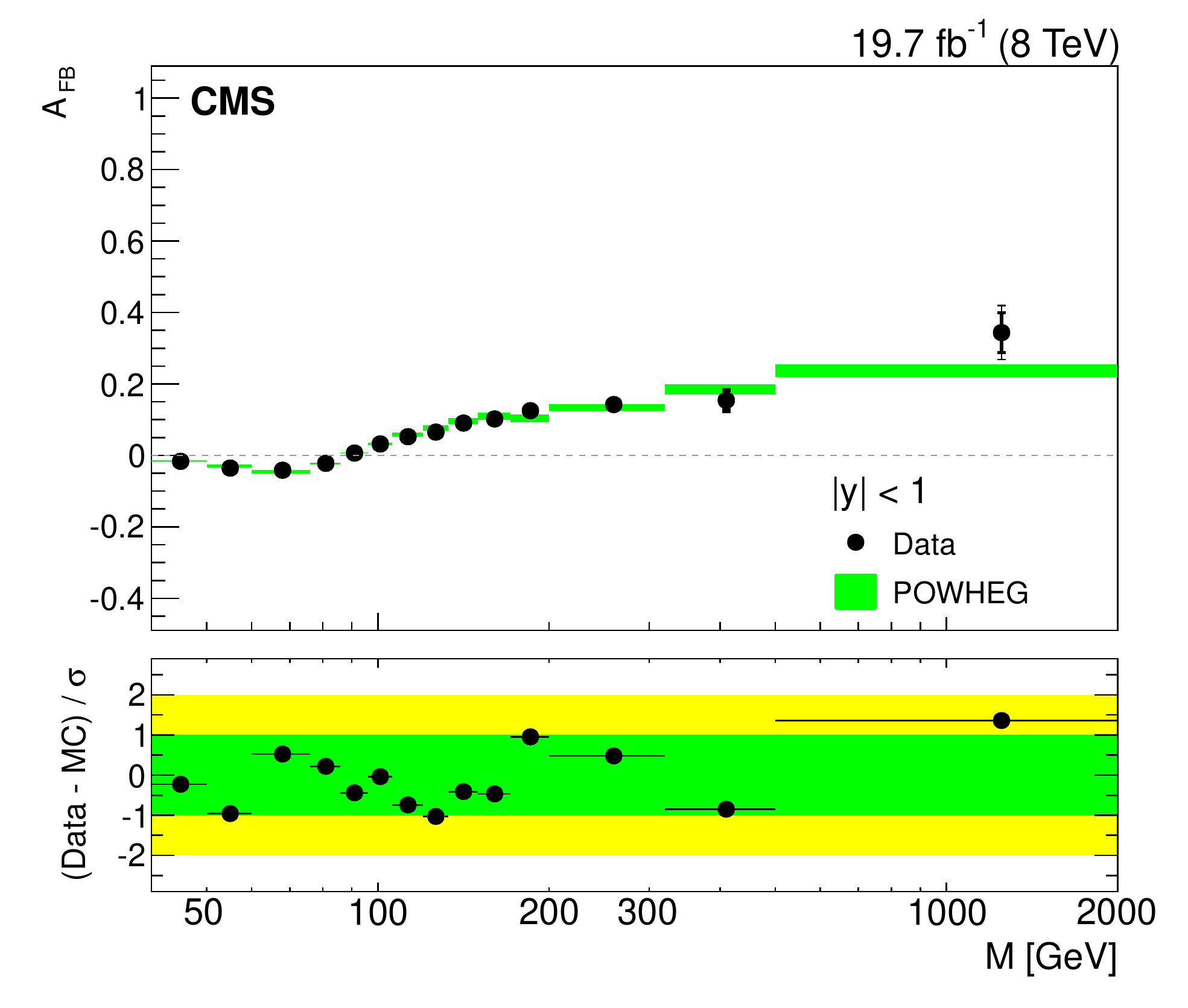}
		\includegraphics[width=0.48\textwidth]{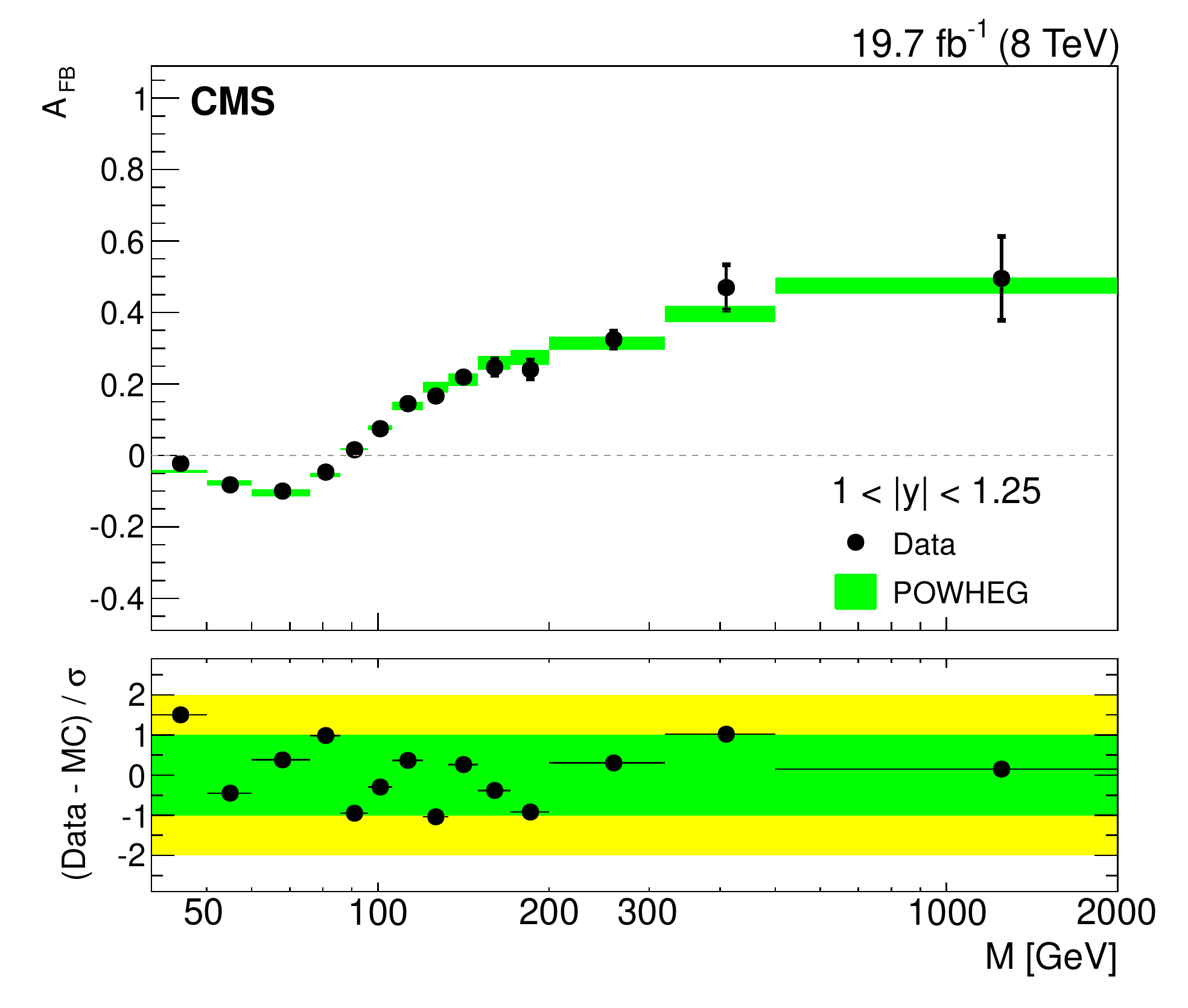}
		\includegraphics[width=0.48\textwidth]{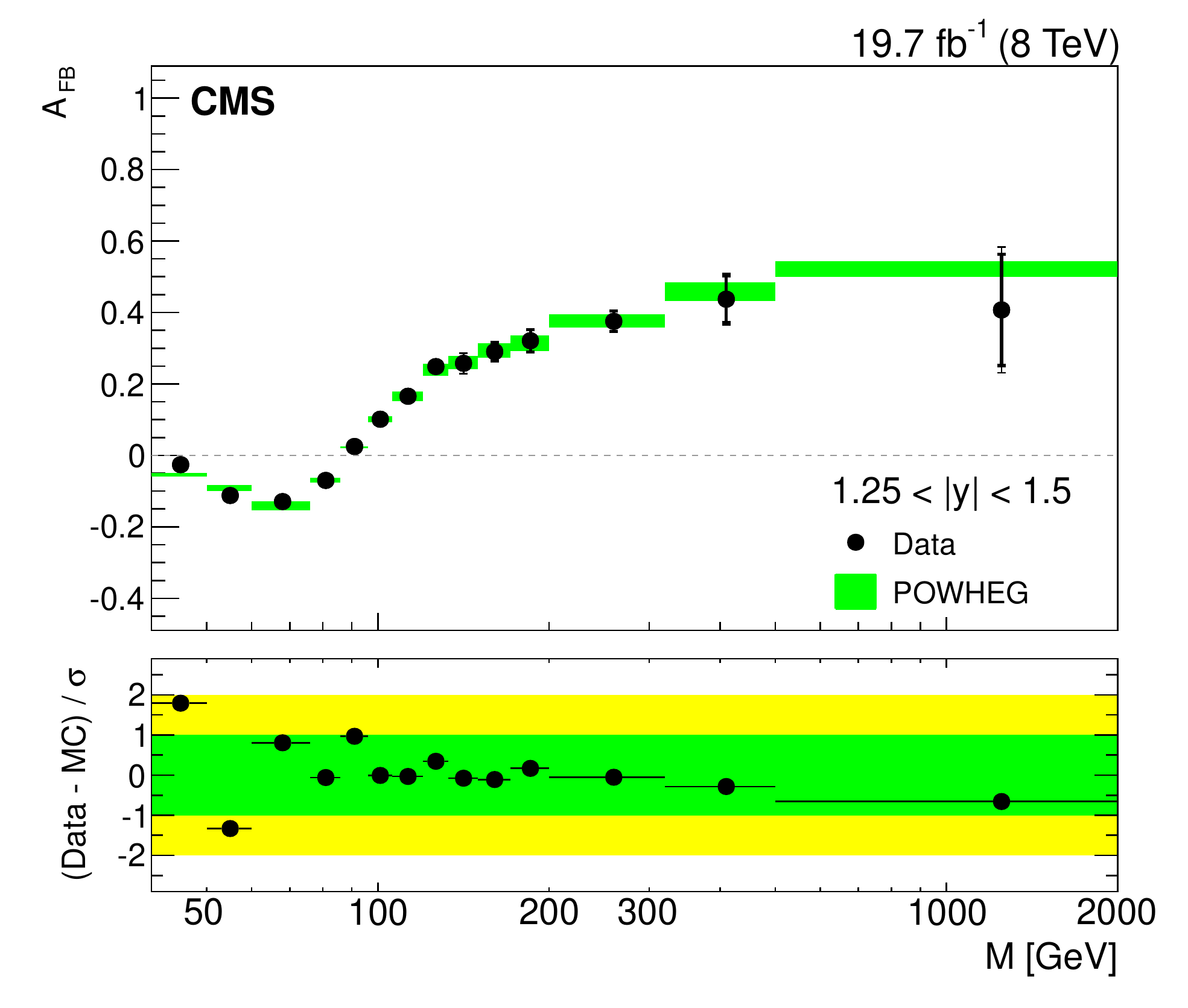}
		\includegraphics[width=0.48\textwidth]{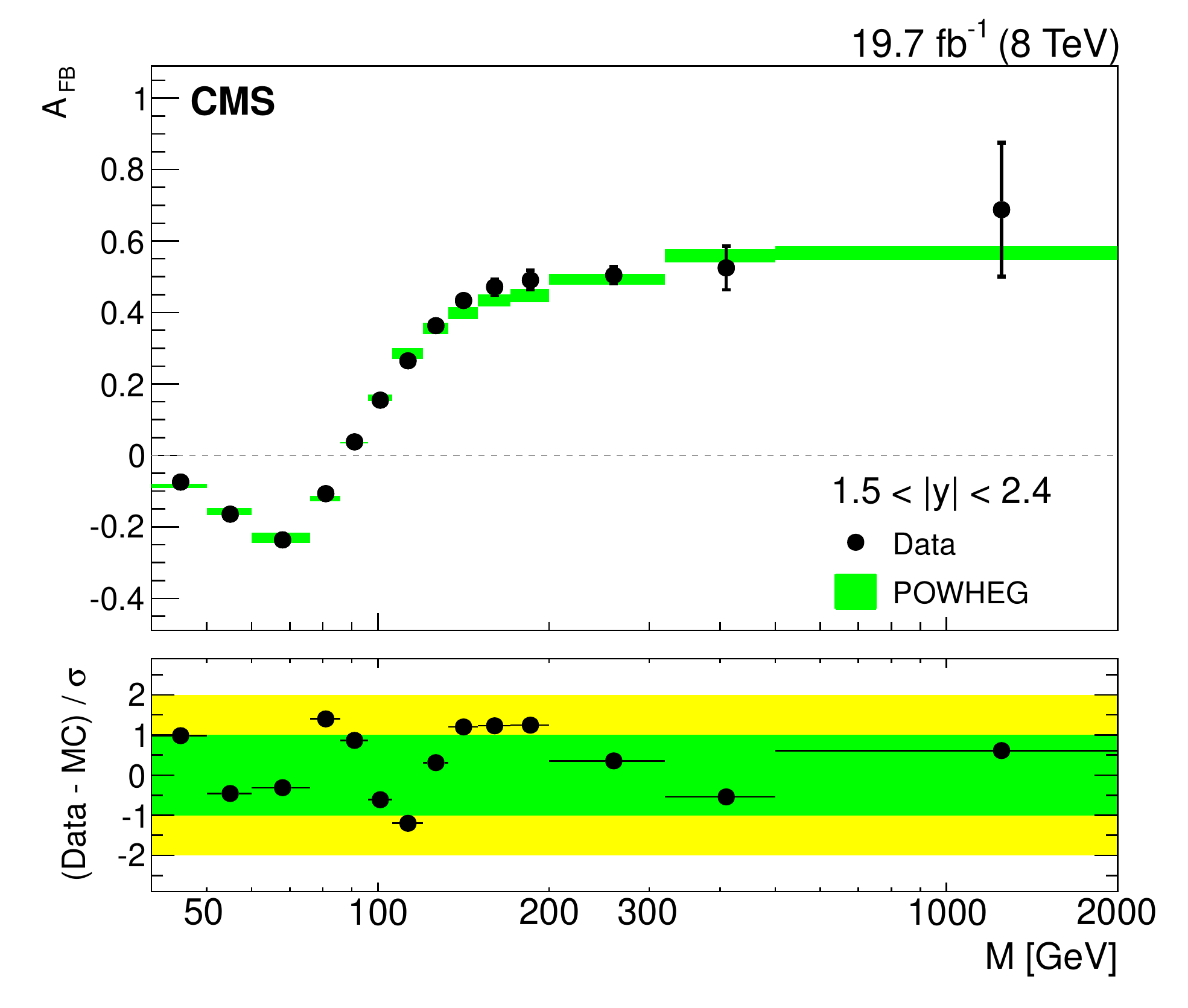}
		\caption{The combined  ($\Pgmp\Pgmm$ and $\Pep\Pem$ ) unfolded \afb distributions  in the  four central rapidity regions. The statistical (thick vertical bar) and statistical plus systematics (thin vertical bar) uncertainties are presented.
The measurements are compared with the prediction of \POWHEG. The total uncertainties (considering the statistical, PDF, and scale uncertainties) in the \POWHEG prediction are shown as shaded bands.
The lower panel in each plot shows the difference of \afb in data and prediction divided by the total uncertainty of data and prediction.}
\label{fig:Afb_combine}
\end{figure*}

Figure~\ref{fig:Afb_combine} shows the combined results for the four central rapidity regions up to 2.4.
The combined result is compared with the \POWHEG (NLO) prediction with CT10 PDFs.
The effective weak mixing angle, $\sin^2\theta^\text{eff}_{\text{lept}}$ = 0.2312, is used for the \POWHEG prediction.
For all rapidity regions, the combined \afbM values are in a good agreement with the \POWHEG prediction.
The uncertainty in the theoretical prediction (\POWHEG) originates from the statistical uncertainty in the MC sample, the uncertainties in the PDFs, and the variations of factorization and renormalization scales (simultaneous variation between values 2$M$, $M$, and $M/2$, with $M$ corresponding to the middle of the invariant mass bin).
 Table~\ref{tab:result} summarizes the combined \afb quantity for each rapidity region.

\begin{figure}[!htb]
	\centering
		\includegraphics[width=0.48\textwidth]{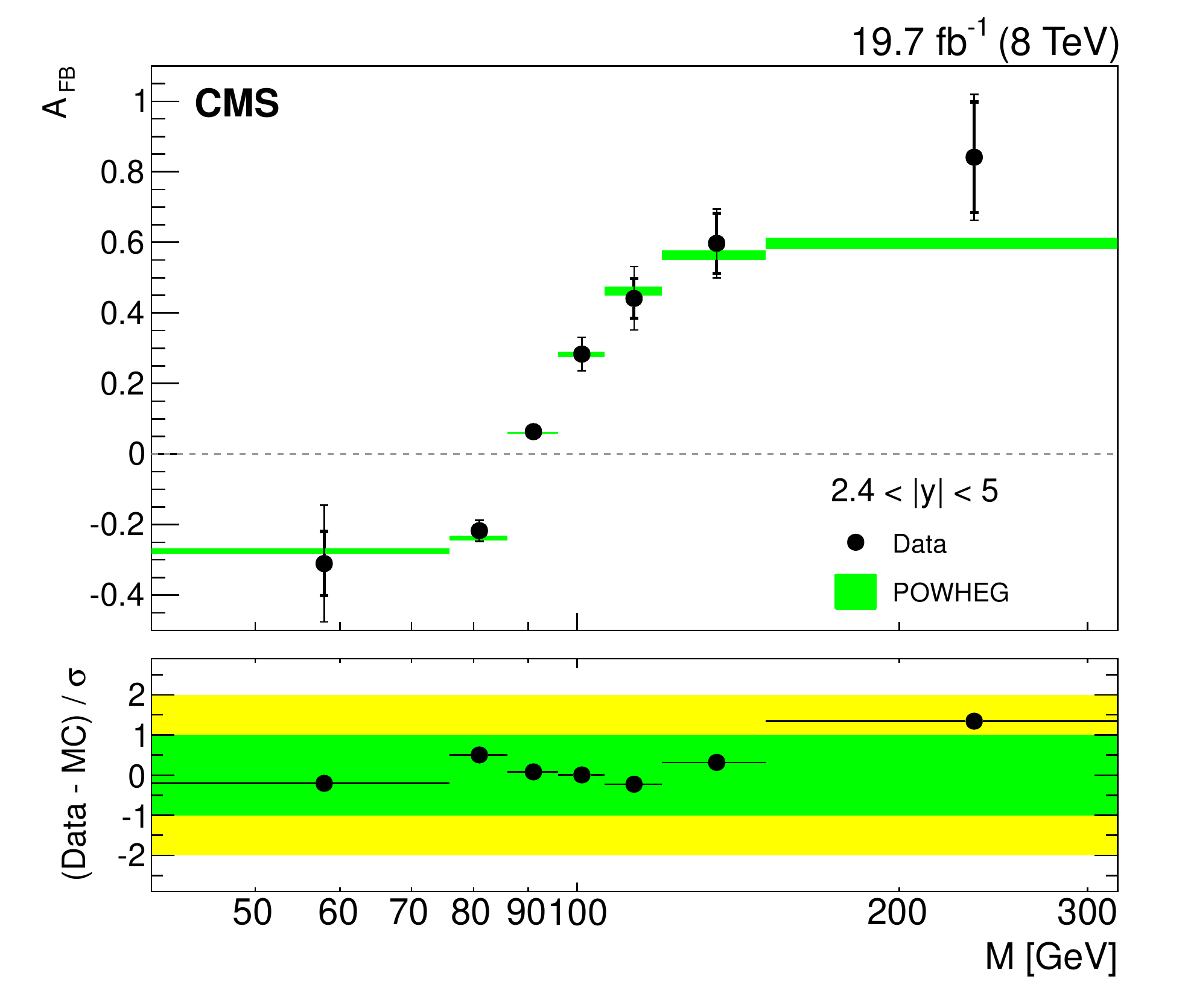}
		\caption{The unfolded \afb distribution for the forward rapidity region ($2.4<\abs{y}<5$) using one central electron ($\abs{\eta}<2.4$) and one HF electron ($3<\abs{\eta}<5$).
 The inner thick vertical bars correspond to the statistical uncertainty and the outer thin vertical bars to the total uncertainties. The measurements are compared with the prediction of \POWHEG.
		The  total uncertainties (considering the statistical, PDF, and scale uncertainties) in the  \POWHEG prediction are shown as shaded bands.
                The lower panel shows the difference of \afb in data and prediction divided by the total uncertainty of data and prediction. }
		\label{fig_afb_hfEle}	
\end{figure}

 The unfolded \afb distribution for the forward rapidity region ($2.4 < \abs{y} < 5$) is shown in Fig.~\ref{fig_afb_hfEle}.
 The forward rapidity region extends the scope of the measurement beyond that of the previous CMS result at $\sqrt{s}=7$\TeV.
Because \afb in the forward rapidity region is diluted less, the measured \afb quantity is closer to the parton-level asymmetry after the unfolding process, than it is in the central rapidity bins.
 The unfolded  \afb($M_{\Pep\Pem}$) for $2.4<\abs{y}<5$ agrees with the \POWHEG predictions.

\begin{table*}[htb]
\centering
\topcaption{The combined ($\Pe\Pe$ and $\mu\mu$) \afb measurements, with statistical
and systematic uncertainties for the four rapidity regions with $\abs{y}<2.4$.
 The \afb quantity for $\Pe\Pe$ events is also shown for $2.4<\abs{y}<5$. }
\label{tab:result}
\newcolumntype{x}{D{,}{\text{--}}{3.4}}
\resizebox{\textwidth}{!}{
\begin{tabular}{xrrrr|xrrrr} \hline
 \multicolumn{1}{c}{$M$ (\GeVns{})}  & \afb (data) & Stat. err &  Syst. err & Tot. err & \multicolumn{1}{c}{$M$ (\GeVns{}) }& \afb (data) & Stat. err  & Syst. err & Tot. err \\ \hline
\multicolumn{5}{c|} {$\abs{y}<1$} & \multicolumn{5}{c}{$1<\abs{y}<1.25$} \\ \hline
40,50    & $-0.0167$ & $0.0049$ & $0.0045$ & $0.0067$  &  40,50    & $-0.0225$ & $0.0108$ & $0.0092$ & $0.0142$   \\	
50,60    & $-0.0355$ & $0.0042$ & $0.0031$ & $0.0052$	&  50,60    & $-0.0825$ & $0.0092$ & $0.0060$ & $0.0110$   \\	
60,76    & $-0.0415$ & $0.0033$ & $0.0031$ & $0.0045$	&  60,76    & $-0.0999$ & $0.0071$ & $0.0044$ & $0.0084$   \\ 	
76,86    & $-0.0221$ & $0.0022$ & $0.0019$ & $0.0029$	&  76,86    & $-0.0468$ & $0.0048$ & $0.0042$ & $0.0064$   \\	
86,96    & $0.0065 $ & $0.0004$ & $0.0003$ & $0.0005$	&  86,96    & $0.0157 $ & $0.0009$ & $0.0005$ & $0.0011$    \\	
96,106   & $0.0320 $ & $0.0020$ & $0.0016$ & $0.0025$	&  96,106   & $0.0747 $ & $0.0046$ & $0.0042$ & $0.0063$   \\ 	
106,120  & $0.0524 $ & $0.0037$ & $0.0024$ & $0.0045$	&  106,120  & $0.1448 $ & $0.0085$ & $0.0029$ & $0.0089$  \\	
120,133  & $0.0652 $ & $0.0065$ & $0.0035$ & $0.0074$	&  120,133  & $0.1663 $ & $0.0152$ & $0.0083$ & $0.0174$  \\	
133,150  & $0.0905 $ & $0.0081$ & $0.0070$ & $0.0108$	&  133,150  & $0.2191 $ & $0.0185$ & $0.0064$ & $0.0195$  \\	
150,171  & $0.1020 $ & $0.0104$ & $0.0075$ & $0.0128$	&  150,171  & $0.2469 $ & $0.0243$ & $0.0123$ & $0.0272$  \\	
171,200  & $0.1251 $ & $0.0129$ & $0.0145$ & $0.0194$	&  171,200  & $0.2401 $ & $0.0272$ & $0.0143$ & $0.0308$  \\	
200,320  & $0.1423 $ & $0.0112$ & $0.0099$ & $0.0149$	&  200,320  & $0.3245 $ & $0.0257$ & $0.0115$ & $0.0282$  \\	
320,500  & $0.1541 $ & $0.0268$ & $0.0195$ & $0.0331$	&  320,500  & $0.4697 $ & $0.0609$ & $0.0302$ & $0.0680$  \\	
500,2000 & $0.3437 $ & $0.0554$ & $0.0514$ & $0.0756$   &  500,2000& $0.4954 $ & $0.1145$ & $0.0400$ & $0.1213$ \\ \hline
\multicolumn{5}{c|} {$1.25<\abs{y}<1.5$} & \multicolumn{5}{c}{$1.5<\abs{y}<2.4$} \\ \hline
40,50   & $-0.0261$ & $0.0114$ & $0.0087$ & $0.0144$   &  40,50   & $-0.0747$ & $0.0073$ & $0.0049$ & $0.0088$  \\	
50,60   & $-0.1122$ & $0.0098$ & $0.0078$ & $0.0125$	&  50,60   & $-0.1645$ & $0.0070$ & $0.0053$ & $0.0088$  \\	
60,76   & $-0.1293$ & $0.0077$ & $0.0039$ & $0.0086$	&  60,76   & $-0.2365$ & $0.0059$ & $0.0052$ & $0.0079$  \\ 	
76,86   & $-0.0700$ & $0.0052$ & $0.0040$ & $0.0065$	&  76,86   & $-0.1071$ & $0.0041$ & $0.0057$ & $0.0070$  \\	
86,96   & $0.0249 $ & $0.0010$ & $0.0007$ & $0.0013$	&  86,96   & $0.0379 $ & $0.0008$ & $0.0009$ & $0.0013$   \\	
96,106  & $0.1012 $ & $0.0051$ & $0.0044$ & $0.0067$	&  96,106  & $0.1546 $ & $0.0041$ & $0.0057$ & $0.0070$  \\ 	
106,120 & $0.1655 $ & $0.0095$ & $0.0045$ & $0.0105$	&  106,120 & $0.2647 $ & $0.0078$ & $0.0047$ & $0.0091$ \\	
120,133 & $0.2485 $ & $0.0169$ & $0.0080$ & $0.0187$	&  120,133 & $0.3630 $ & $0.0141$ & $0.0068$ & $0.0156$ \\	
133,150 & $0.2576 $ & $0.0210$ & $0.0197$ & $0.0287$	&  133,150 & $0.4334 $ & $0.0179$ & $0.0129$ & $0.0221$ \\	
150,171 & $0.2903 $ & $0.0259$ & $0.0103$ & $0.0279$	&  150,171 & $0.4713 $ & $0.0230$ & $0.0083$ & $0.0245$ \\	
171,200 & $0.3209 $ & $0.0315$ & $0.0112$ & $0.0335$	&  171,200 & $0.4906 $ & $0.0276$ & $0.0095$ & $0.0292$ \\	
200,320 & $0.3752 $ & $0.0286$ & $0.0114$ & $0.0308$	&  200,320 & $0.5042 $ & $0.0244$ & $0.0092$ & $0.0261$ \\	
320,500 & $0.4372 $ & $0.0655$ & $0.0287$ & $0.0715$	&  320,500 & $0.5248 $ & $0.0610$ & $0.0131$ & $0.0624$ \\	
500,2000& $0.4071 $ & $0.1556$ & $0.0824$ & $0.1761$   &  500,2000& $0.6878 $ & $0.1862$ & $0.0413$ & $0.1907$\\ \hline
\multicolumn{5}{c|} {$2.4<\abs{y}<5$ ($\Pe\Pe$ only)}   &  \\ \cline{1-5}
40,76   & $-0.3104$ & $0.0912$ & $0.1378$ & $0.1652$   &  \\ 	
76,86   & $-0.2174$ & $0.0214$ & $0.0210$ & $0.0300$   &  \\ 	
86,96   & $0.0635 $ & $0.0060$ & $0.0146$ & $0.0158$    & \\ 	
96,106  & $0.2834 $ & $0.0183$ & $0.0439$ & $0.0475$    & \\ 	
106,120 & $0.4412 $ & $0.0567$ & $0.0696$ & $0.0898$  & \\ 	
120,150 & $0.5972 $ & $0.0851$ & $0.0476$ & $0.0975$  & \\ 	
150,320 & $0.8412 $ & $0.1567$ & $0.0851$ & $0.1783$  & \\  \cline{1-5}
\end{tabular}}
\end{table*}

\section{Summary}

We report a measurement of the forward-backward asymmetry of oppositely charged $\mu\mu$ and $\Pe\Pe$ pairs produced via a $\Z/\gamma^*$ boson exchange at $\sqrt{s} = 8$\TeV with a data sample corresponding to an integrated luminosity of 19.7\fbinv.
The \afb measurement is performed as a function of the dilepton invariant mass between 40\GeV and 2\TeV for  $\mu\mu$ and $\Pe\Pe$ events in 4 dilepton rapidity bins up to $\abs{y} = 2.4$. For $\Pe\Pe$ events with $2.4 < \abs{y} < 5$, the \afb measurement is performed for dielectron masses between 40 and 320\GeV. The large data sample collected at 8\TeV extends the measurement of \afb in the high mass region compared to previous results. The final \afb values are corrected for detector resolution, acceptance, and final state radiation effects. The measurements of \afbM are consistent with standard model predictions.

\begin{acknowledgments}
We congratulate our colleagues in the CERN accelerator departments for the excellent performance of the LHC and thank the technical and administrative staffs at CERN and at other CMS institutes for their contributions to the success of the CMS effort. In addition, we gratefully acknowledge the computing centers and personnel of the Worldwide LHC Computing Grid for delivering so effectively the computing infrastructure essential to our analyses. Finally, we acknowledge the enduring support for the construction and operation of the LHC and the CMS detector provided by the following funding agencies: BMWFW and FWF (Austria); FNRS and FWO (Belgium); CNPq, CAPES, FAPERJ, and FAPESP (Brazil); MES (Bulgaria); CERN; CAS, MoST, and NSFC (China); COLCIENCIAS (Colombia); MSES and CSF (Croatia); RPF (Cyprus); MoER, ERC IUT and ERDF (Estonia); Academy of Finland, MEC, and HIP (Finland); CEA and CNRS/IN2P3 (France); BMBF, DFG, and HGF (Germany); GSRT (Greece); OTKA and NIH (Hungary); DAE and DST (India); IPM (Iran); SFI (Ireland); INFN (Italy); MSIP and NRF (Republic of Korea); LAS (Lithuania); MOE and UM (Malaysia); CINVESTAV, CONACYT, SEP, and UASLP-FAI (Mexico); MBIE (New Zealand); PAEC (Pakistan); MSHE and NSC (Poland); FCT (Portugal); JINR (Dubna); MON, RosAtom, RAS and RFBR (Russia); MESTD (Serbia); SEIDI and CPAN (Spain); Swiss Funding Agencies (Switzerland); MST (Taipei); ThEPCenter, IPST, STAR and NSTDA (Thailand); TUBITAK and TAEK (Turkey); NASU and SFFR (Ukraine); STFC (United Kingdom); DOE and NSF (USA).

Individuals have received support from the Marie-Curie program and the European Research Council and EPLANET (European Union); the Leventis Foundation; the A. P. Sloan Foundation; the Alexander von Humboldt Foundation; the Belgian Federal Science Policy Office; the Fonds pour la Formation \`a la Recherche dans l'Industrie et dans l'Agriculture (FRIA-Belgium); the Agentschap voor Innovatie door Wetenschap en Technologie (IWT-Belgium); the Ministry of Education, Youth and Sports (MEYS) of the Czech Republic; the Council of Science and Industrial Research, India; the HOMING PLUS program of the Foundation for Polish Science, cofinanced from European Union, Regional Development Fund; the OPUS program of the National Science Center (Poland); the Compagnia di San Paolo (Torino); MIUR project 20108T4XTM (Italy); the Thalis and Aristeia programs cofinanced by EU-ESF and the Greek NSRF; the National Priorities Research Program by Qatar National Research Fund; the Rachadapisek Sompot Fund for Postdoctoral Fellowship, Chulalongkorn University (Thailand); the Chulalongkorn Academic into Its 2nd Century Project Advancement Project (Thailand); and the Welch Foundation, contract C-1845.
\end{acknowledgments}

\clearpage
\bibliography{auto_generated}

\cleardoublepage \appendix\section{The CMS Collaboration \label{app:collab}}\begin{sloppypar}\hyphenpenalty=5000\widowpenalty=500\clubpenalty=5000\textbf{Yerevan Physics Institute,  Yerevan,  Armenia}\\*[0pt]
V.~Khachatryan, A.M.~Sirunyan, A.~Tumasyan
\vskip\cmsinstskip
\textbf{Institut f\"{u}r Hochenergiephysik der OeAW,  Wien,  Austria}\\*[0pt]
W.~Adam, E.~Asilar, T.~Bergauer, J.~Brandstetter, E.~Brondolin, M.~Dragicevic, J.~Er\"{o}, M.~Flechl, M.~Friedl, R.~Fr\"{u}hwirth\cmsAuthorMark{1}, V.M.~Ghete, C.~Hartl, N.~H\"{o}rmann, J.~Hrubec, M.~Jeitler\cmsAuthorMark{1}, V.~Kn\"{u}nz, A.~K\"{o}nig, M.~Krammer\cmsAuthorMark{1}, I.~Kr\"{a}tschmer, D.~Liko, T.~Matsushita, I.~Mikulec, D.~Rabady\cmsAuthorMark{2}, B.~Rahbaran, H.~Rohringer, J.~Schieck\cmsAuthorMark{1}, R.~Sch\"{o}fbeck, J.~Strauss, W.~Treberer-Treberspurg, W.~Waltenberger, C.-E.~Wulz\cmsAuthorMark{1}
% ...
\vskip\cmsinstskip
% ...
1:~~Also at Vienna University of Technology, Vienna, Austria\\
2:~~Also at CERN, European Organization for Nuclear Research, Geneva, Switzerland\\
% ...

\end{sloppypar}
\end{document}